\documentstyle[12pt]{article}

\title{Heterogeneous decay of metastable phase on various centers - 1}
\author{V.Kurasov \\
St.Petersburg State University, \\
Department of Mathematical and Computational Physics}

\date{Victor.Kurasov@pobox.spbu.ru }

\begin{document}
\maketitle
\begin{abstract}

A system of a metastable phase with several sorts of
heterogeneous centers is
considered.
An analytical theory for the process of
decay in such a system has been constructed.
The free energy of  formation of a critical embryo  is assumed to be known in
the
 macroscopic approach. At first all
asymptotic cases are investigated and then a general intermediate solution is
suggested.
All approximate transformations are accomplished with the corresponding
numerical
estimates and  analytical justifications.
This is the first part of the manuscript.  the second part follows
the first one in the e-print archive.

\end{abstract}

\pagebreak

\section{Introduction}

The kinetics of condensation on various
centers which simultaneously exist in the system will be constructed here.
The case of   condensation of a supersaturated vapor into  a state
of  liquid droplets seems to be the simplest case among the first order
phase transitions. Traditionally this case  is considered as a model to
introduce some new theoretical  constructions in  description of the
first order phase transitions. Ordinary the process of nucleation occurs
on dust particles   or on heterogeneous centers. A case of heterogeneous
condensation was a first object in  regular investigations of the
first order phase transitions made by Wilson \cite{wilson}. A first
technical device constructed by Wilson, i.e. the famous Wilson chamber
was based on the effect of heterogeneous condensation (see also \cite{wilson}).
A problem of the chamber purification is one of
actual problems in experimental observations of  the homogeneous condensation
and the heterogeneous way of nucleation is the most usual way of the
droplets formation in  nature.
So, the case of heterogeneous condensation which will be investigated
is very wide spread in nature.

The case of condensation is the most well investigated example of the
first order phase transitions.
This leading role of the case  of
condensation was outlined by  creation  of the  classical theory of
 nucleation by Volmer \cite{Volmer}, Becker and Doering \cite{Becker},
Zeldovitch
\cite{Zeld}  which gave for the first time
an expression for the rate of nucleation (i.e. for the rate of an
appearance
of new droplets), which will be
an elementary brick in further constructions.
The careful analysis of the classical  theory  leads to  great number
of the publications with   various reconsiderations of  classical
expressions. Among them  one has to  notice the  account of the internal
degrees of freedom in the embryo made
by Lothe and Pound \cite{Lothe}, another modifications
made by Reiss, Cohen and Katz \cite{RCK}, Reiss \cite{Reiss1}, Fisher \cite{Fisher}.
The application of the density functional theory  to the first order phase
transition made by R.Evans and D.Oxtoby \cite{EvOxt}, D.Oxtoby and
D.Zeng \cite{ZengOxt} allowed to put  the microscopic
(mesoscopic) models for the  condensated substance as the base for an
expression for the free energy of the critical embryo and for the rate
of nucleation.
  In the  publications
of Oxtoby and Talanquer \cite{OxtTal},
Reiss, Tabazadech and Talbot
\cite{RTT}, Reiss, Ellerby and Weakliem \cite{REW1} - \cite{REW4}
   the role
of the environment around a droplet is carefully analysed.  It
 is          equivalent  to some specific  choice of
the statistical ensemble.

Nevertheless one has to notice that there is no perfect coincidence between
 concrete theoretical predictions\footnote{Certainly, one can not
check directly some general recipes
as to calculate the objects like  statistical
sums.} and  experimental results.
So, this question is still opened in the modern state of investigations.
But the
mentioned expressions for the nucleation rate reproduce
a qualitative behavior of  experimental results quite adequately.
Moreover, a relative
deviation between  theoretical predictions
and  experimental results is very smooth function of parameters.

The mentioned publications  allowed to start an  investigation of  the global kinetics
 of heterogeneous condensation.
The qualitative description of the global evolution during
the first order phase transition was initiated
by Wakeshima \cite{Wak}
who considered  time lags (characteristic times) for  condensation.
The characteristic time of  formation of the droplets spectrum
was necessary to ensure a correct experimental definition of the stationary
rate of nucleation.
After \cite{Wak} the interest for the description of the global evolution
was growing continuously.

The nature and the content of aerosol or heterogeneous
particles are so various that  different approaches to describe
the process of condensation are required.
Speaking about the aerosol particles
one supposes that these particles are already the supercritical objects
of a liquid phase. These objects are growing irreversibly and regularly
in time while the objects of a liquid phase formed on  heterogeneous centers
have to overcome the activation barrier. The latter can be done fluctuationally
and
the process of the supercritical embryo appearance
 resembles the homogeneous nucleation, but the height of activation
barrier is smaller than in the homogeneous case.

Here the supercritical objects of a liquid phase (may be with  a heterogeneous
center inside) will be called as droplets while the objects of a liquid
phase with an arbitrary
size      and nature will be called as "embryos" or as "particles of a new
phase".
Very often there is no need to consider the process of the droplets formation
because in the system
there is  a sufficient quiantity of aerosol in the system. This case
is extracted by the evident simplification that the total number of droplets
is already known. Sometimes this case is described as a barrierless formation
of droplets. The description of this situation one can see in
 \cite{Clement-Royal} where this case was completely investigated.

When the activation barrier exists then the total number of droplets is unknown
and one has to determine a number of droplets and their size spectrum
and has to solve a complex
non-linear problem.
To solve this problem numerical calculations
were presented in \cite{Warren-Seinfield-JCIS}, \cite{Warren-Seinfield-AST3}.
The numerical method of calculations presented in  \cite{Warren-Seinfield-AST3}
allows to to establish in \cite{Warren-Seinfield-JCIS}
some dimensionless combinations which  essentially  simplifies  the
numerical
procedure. This simplification
gives a way  to fulfill some rather complex numerical calculations.

The next step in the development of the global evolution description was
the sectional model  \cite{Warren-Seinfield-AST4} which  simplifies
calculations one more time and allows to take into account both nucleation
and coagulation.
In this  publication the process of coagulation isn't described
because the probability of this process is very low and it can be observed
only  a long time after the end of nucleation, i.e. the end of the process
of intensive formation of super-critical embryos.

The main difficulty of the nucleation kinetics is a necessity to take into account
the exhaustion of a vapor phase due to a vapor consumption
by droplets. This exhaustion diminishes the nucleation rate for new
imaginary droplets.
Sometimes this influence isn't essential and it is shown in
\cite{Barrett-Badwin}
that this situation is rather wide spread in  laminar  flows. In
\cite{Barrett-Badwin} a  theory for this case was given and some methods
to describe the global evolution were presented.
The great importance of the
problem of the vapor exhaustion around the droplet was stressed
by H.Reiss  in \cite{Reiss} where the stationary profiles around droplets
were obtained.

To take this influence into account one has to solve a non-linear problem. The
simplest way
to analyze this problem
is to act in frames of some characteristic values of time and space.
Namely
this way was chosen
in \cite{Barrett}.

An effect of a heterogeneous centers exhaustion also diminishes the intensity
of the new droplets appearance. This effect also has to be taken into
account. One has to solve this problem in   a self consistent way.
The theory for the condensation on one type of heterogeneous centers
can be found in \cite{PhysRevE}, where the approximate self consistent solution
of the balance equations is given.

Now one has to specify the situation considered in this
publication.
One has to mention that ordinary in the system there exist several types
of heterogeneous centers of  different nature.
These centers can have
different values  of the activation barriers under the given value of
metastability in the system. Sometimes the difference
between centers with neighbour characteristics  is so small that
one can speak about the quasi-continuous activity of the band of heterogeneous
centers. But sometimes the difference in  barriers heights between centers with neighbor
characteristics leads to their different kinetic behavior.

To show the variety of situations  two examples will be given.
In the process of condensation on the ions the free energy
of the critical embryo depends  on the sign of  electric charge.
As it is shown
in \cite{Rus} the free energy $F$ of an embryo with $\nu$ molecules inside
can be presented as
\begin{equation} \label{1}
F=-b\nu+a\nu^{2/3}+c_{1}\nu^{1/3}+(c_{2}+c_{3})\nu^{-1/3} + c_{0} \ln \nu +
const \ \ .
\end{equation}
Here and later all energy-like
values are expressed in  units of the mean thermal energy; $
 a, b, c_{0}, c_{1}, c_{2}, c_{3} $
are some parameters.
This formula is valid  for the embryo in the state of internal
equilibrium, i.e. near the value $\nu_c$  of a number of molecules
in the critical embryo (corresponding to the maximum of $F$) and
near the value $\nu_e$ of a number of molecules
in the equilibrium embryo (corresponding to the minimum of $F$).
All values at $\nu=\nu_c$ will be marked by the lower index $c$ and
all values at $\nu=\nu_e$ will be marked by the lower index $e$.
It is necessary to notice that  $a, b, c_{0},
 c_{1}, c_{2}$  don't depend
on the sign of the charge $q$  and a value of $c_{3}$ is proportional to
$q$.
Since $\nu_{e} < \nu_{c}$ then the height of activation barrier $\Delta
F=F_c - F_e$    depends on the sign
of $q$.
So, in   presence of
radiation one has two sorts of  centers
(positive and negative) with  different heights of  activation barriers,
i.e. with  different activities of  heterogeneous centers.

The spectrum of the sizes of heterogenenous centers as the solid
balls  with a weak interaction
between the heterogeneous center and a liquid phase
leads to the
spectrum of activities of  heterogeneous centers. In the
simple model with the passive nuclei  one has
 to add to the surface term the number of the molecules imaginary
contained in the volume occupied by the nuclei. The free energy of the
critical embryo is given by
\begin{equation}
F = - b \nu + a
(\nu+\frac{4\pi r^3} {3 v_l})^{2/3} + const \ \ ,
\end{equation}
where $r$ is the
radius of a heterogeneous center,
$v_l$ is the volume occupied by a molecule of a
 substance in a liquid phase and  $F_e \ = \ const$.
  A continuous spectrum of sizes of
heterogeneous centers  initiates a continuous  spectrum       of
the activation barriers heights.

As the result   two different situations can take
place - the situation with several types of centers and the
situation  with continuous variation of the properties of centers.
Any spectrum of  activation barrier heights can be split
into  several continuous parts (may be also like $\delta$-functions) which
will be considered as the "types" of  heterogeneous centers.

Now one has to specify external conditions.
Kinetic description of the condensation process implies   external
conditions to be known. There are two characteristic types of external
conditions. The first one is the external conditions of the decay type -
until some moment the system is in the stable state and then rather rapidly
due to the action of external forces the initial phase becomes metastable.
Then the external action stops.
 This type of conditions is deeply
connected with  typical conditions in the   chamber experiments.
Investigations of the condensation kinetics \cite{Wak} were started
in this situation.

The second typical conditions are conditions of the dynamic type. The
action of external conditions gradually creates metastability
but a  rapid growth of a  vapor consumption by droplets at some
moment begin to compensate  an action of external conditions, which doesn't stop.
The rate of
a vapor consumption grows so rapidly that  the
power of metastability  falls  and the process of nucleation stops.
 This situation is described in \cite{PhysRevE}
for a system with
one type of heterogeneous centers.

Why the situation of decay on several types of heterogeneous
centers requires a special consideration?
For the condensation on one type of heterogeneous centers one can see
a certain  analogy between  two mentioned
 types of external conditions. This is explained
by the evident notation that in both situations the intensive formation
of new droplets occurs near some
characteristic value of the power of metastability\footnote{This
value is determined by external conditions.}.
During the
nucleation on different sorts of heterogeneous centers in the situation
of decay the characteristic value of the power of metastability is
one and the same for all types of centers
and in the situation of dynamic external conditions the characteristic
value of metastability is determined separately for every type of centers.
This explains why one has to use different methods for  different
specific situations  of heterogeneous condensation on many types of centers.
In the current publication
 the situation of decay
for heterogeneous
centers with arbitrary
activities
 will be analyzed.

One has to study  the period of the intensive formation of the droplets, i.e.
the nucleation period.
The further evolution is described analogously to the condensation on
one type of centers.

The following physical assumptions to formulate the
model will be used: the
 thermodynamic description of the critical embryo; the
 random homogeneous space distribution of
heterogeneous centers;                          the
 homogeneous
external conditions for    temperature and   pressure;
the   high   (in comparison with one thermal unit)
activation barrier\footnote{The theory without the heterogeneous activation
barrier is  much more simple.};                    the
 absence of  thermal effects.
One can see that these assumptions are rather natural.  The unit volume is considered.

According to  \cite{PhysicaA} the condensation equations in the
general conditions  are analogous to the case
of the free molecular consumption of the vapor and  namely this
case will be studied.

A homogeneous character of a distribution of heterogeneous centers
implies that in a characteristic space region  there is
a big quantity of heterogeneous centers of every type\footnote{For the
quasi continuous spectrum of activities the different type means that the
difference of the activation barrier height is greater than one thermal
unit.}. The characteristic space size of this region is determined as a
mean distance of the diffusion  relaxation during the whole nucleation
period. For the pure free molecular regime of droplets growth
the characteristic size is the size of a system.

The total number of the heterogeneous centers is assumed to be constant in
time. This is also quite natural because the period of nucleation is relatively
short in comparison with other characteristic times of  the condensation
process (except the time of relaxation to the stationary state in the near-critical
region).

Since the most interesting characteristics of this process are the
 numbers of
 droplets on the different types  of centers,
the accuracy of the theory will be estimated
by  an error in this value.

\section{Kinetic equations}

 The total number of  heterogeneous centers will be marked by
$\eta_{tot\ i}$, where $i$ corresponds to some sort of the heterogeneous centers.
The absence of this index points that
the formula is valid for an arbitrary sort of  heterogeneous centers.
The numbers of
the free heterogeneous centers which
 aren't occupied by
the super critical embryos are marked by the value of $\eta_{i}$.

One can define the neighbor sorts of heterogeneous centers as the centers
with neighbor values  of the activation barriers heights.
When the difference in the activation barriers heights for the neighbor sorts is many
times less than one thermal unit one can speak about the quasi continuous
spectrum of activities of heterogeneous centers. The latter doesn't mean
that
the total variation of the activation barriers heights is less than one
thermal unit.

It is convenient to introduce the characteristic
of "activity".
One can define "an activity of a heterogeneous center" as some parameter $w$
which
is proportional to a deviation in  the height of the activation barrier
\begin{equation} \label{**}
\Delta F (w) = \Delta F \mid_{w=0} - \lambda w
\end{equation}
with some positive parameter $\lambda$.
The base of decomposition $(w=0)$  is now an arbitrary
value and  will be chosen later to simplify  formulas.

A total number of  heterogeneous centers
with a given activity $w$ will be marked by $\eta_{tot}(w)$. Naturally
$\eta_{tot}(w)$ is   a smooth function of $w$. Then the neighbor
sorts of heterogeneous centers have  neighbor rates of the droplets
formation\footnote{The sort of the droplets  means the sort of the
heterogeneous centers.}.

          A density
of the molecules in the equilibrium vapor is marked
by $ n_{\infty} $, a
density
of the molecules in the real vapor is marked by  $ n$.
 A power of the metastability will be described by the
 supersaturation
$$ \zeta = \frac{ n - n_{\infty} }{ n_{\infty} } \ \ \ .$$

Every droplet can be described
by the linear size $ \rho = \nu^{1/3} $.
Due to the free-molecular regime of the vapor consumption
\begin{equation} \label{grr}
 \frac{d\rho}{dt} = \zeta \alpha \tau^{-1}                         \ \ ,
\end{equation}
where \( \alpha \) is the condensation coefficient and \( \tau \) is
some characteristic
collision time easily obtained from the gas kinetic theory.

The following statements can be analytically proved:
\begin{itemize}
\item
 The main role in the vapor consumption during nucleation
is played by the super critical embryos, i.e. by the droplets.
\item
 The quasi stationary approximation for the nucleation rate is valid during the
nucleation period.
\end{itemize}
Justification  of the second statement uses the estimate for the times
 \( t^{s}_{i} \)
of the establishing of the stationary  state
in the near-critical region  which can be found in \cite{Zeld}
(for  heterogeneous case  the consideration is  similar).
They have to be many times less than  duration
of  nucleation period.
There may exist     some
 big  times $t^{s}_{i}$. They correspond to  big values
of $\nu_c$.
It means that $\Delta_i F$ are huge and these
sorts of  heterogeneous centers are passive and out of intensive nucleation.

The characteristic time $t_*$
necessary as the base for  decompositions
will be the time of  beginning of
nucleation. The values at the moment $t_*$ will be marked
by the lower index $*$. One can choose the zero point of the time axis
as $t_*$ and  introduce the frontal size \( z \)
according to
\begin{equation}
\label{2}
z = \int_{0}^{t} \zeta \alpha \tau^{-1} dt'                   \ \ .
\end{equation}
Until the coalescence eq.
 (\ref{2}) ensures the growth of \( z \) in time and can be
inverted
\begin{equation} \label{3}
t(z) =\int_{0}^{z}  \tau \alpha^{-1} \frac{dx}{\zeta(x)} \ \ .
\end{equation}
All functions of time become the functions of
$z$ and the relative size $ x=z-\rho $ can be introduced.
During the whole evolution the droplet has one and the same
value of the variable $ x $.
Considering the value $ t(x) $ as the moment when the droplet with a
given $x$ has
been formed (as a droplet, i.e. it begins to grow irreversibly)
one can consider all functions of
time as  functions
of $ x $ or $z$.
The variables $x$ and $z$ become equivalent.
Hence, one can see that the kinetic equation in the supercritical region
is reduced to the
fact that every droplet keeps the constant value of $x$. To reconstruct the
picture
of  evolution one must establish the dependencies $t(z)$ and $\zeta(x)$.

The argument $ \infty $ will mark the total values of the characteristics
formed during the whole nucleation  process.
Practically immediately after creation of metatsability   the value of the
supersaturation falls down to
\begin{equation} \label{4}
\Phi_{*} = \zeta (0) - \frac{
\sum_{i} \eta_{tot\ i}\nu_{e\ i}}{n_{\infty}} \ \ .
\end{equation}
During the period of the essential formation of the droplets  one
can assume that the value $ \nu_{e\ i} $ is  constant   and take
it at $ \zeta = \Phi_{*} $.
In initial approximation $ \nu_{e\ i}$ can be taken at $ \zeta = \zeta(0) $.

One can analytically prove
during the period of nucleation on some given sort
for the variation of the  supersaturation
$$
\mid \zeta - \zeta_* \mid \leq \frac{\Phi_*}{\Gamma_i} \ \ ,
$$
where
\begin{equation}                   \label{6}
\Gamma_{i} = -\Phi_{*}
\frac{d \Delta_{i} F(\zeta)}{d \zeta }  \mid_{\zeta=\Phi_{*}} \ \ .
\end{equation}
The same is valid for all sorts of centers when  $\Gamma_i$
is substituted by $min_i \Gamma_i$.

Let $ f_{*\ i} $ be the amplitude value of the
distribution $ f_i $ of sizes of  heterogeneously
 formed droplets
 measured in units of $ n_{\infty} $.
According to the second statement the value of distribution equals to
the stationary distribution which is  the stationary rate
of nucleation divided by the droplets rate of growth
and by the value
of $n_{\infty}$. Then the stationary distribution $f_i$  can be easily
calculated by the following known formulas \cite{Zeld}:
\begin{equation} \label{7}
\frac{J_i \tau}{\alpha \zeta n_{\infty}} =
f_i =\frac{ W^{+}_{c} \exp(-\Delta_i F) \tau}
{ n_{\infty} \pi^{1/2} \Delta_{e\ i} \nu  \Delta_{c\ i} \nu  \zeta  \alpha }
\eta_i
\equiv f_{\zeta \ i } \eta_{i} \ \ ,
\end{equation}
where $W^{+}$ is the number of the molecules absorbed by the
 embryo  in a unit
of time, $\Delta_{e} \nu$  is the characteristic width of
the equilibrium distribution
$$
\Delta_{e\ i} \nu = \sum_{\nu=1}^{\nu=(\nu_{c}+\nu_{e})/2}\exp(-F(\nu))
$$
and $\Delta_{c\ i} \nu$ is the half-width of the near-critical
region\footnote{Here  the steepens descent approximation is used.}
$$
\Delta_{c} \nu =
\frac{2^{1/2}}{\mid \frac{\partial^{2} F }{\partial
\nu^{2}}\mid^{1/2}_{\nu=\nu_{c}}} \ \ .
$$

For the majority of  types of  heterogeneous centers
 the following approximations  of the nucleation rates \( J_{i} \)
 are valid during the nucleation period
\begin{equation}          \label{5}
J_{i} = J_{i}( \eta_{tot\ i} , \Phi_{*} )
\exp ( \Gamma_{i} \frac{ ( \zeta - \Phi_{*} ) }
{ \Phi_{*} } )
\frac{\eta_i}
{\eta_{tot\ i}}  \ \ .
\end{equation}
The validity of these approximations can be easily justified
for  monotonous interaction between the center and the molecules of the
 substance which is
 weaker or equal than the function reciprocal to a space distance.
For the centers with another  interaction this approximation has to be checked
directly.

 The total number of
molecules in droplets formed on the sort "i" is
marked by \( n_{\infty} g_{i} \).
To simplify the formulas $ \theta_{i} =
\eta_{i}/\eta_{tot\ i }  $ will be used.

Using the conservation laws for the  heterogeneous centers
and for the
  molecules
of the substance one can get  for \( g_{i},  \theta_{i} \) the following
equations\footnote{Here the first statement is used.}
\begin{equation}\label{8}
g_{i} = f_{*\ i}  \int_{0}^{z} (z-x)^{3}
\exp ( -\Gamma_{i} \frac{ \sum_{j}g_{j}  }
{ \Phi_{*} } )
\theta_{i} dx
\equiv
G_{i}(\sum_{j}g_{j}, \theta_{i} ) \ \ ,
\end{equation}
\begin{equation}\label{9}
\theta_{i} = \exp ( - f_{*\ i} \frac{n_{\infty}}{\eta_{tot\ i}} \int_{0}^{z}
\exp ( - \Gamma_{i} \frac{ \sum_{j}g_{j}  }
{ \Phi_{*} } ) dx )
\equiv
S_i( \sum_{j}g_{j}) \ \ ,
\end{equation}
where $f_{*\ i} = J_{i}(\eta_{tot\ i},\Phi_{*})\tau /
\Phi_{*} \alpha n_{\infty}$.
These equations form the closed system of the condensation
 equations. This system will be the subject of  investigation.

For the quasi-continuous case one can use the value of activity $w$
as an argument instead
of the lower index  for the sort of heterogeneous centers
and  get more general approximation which
covers also the dependence over activity
\begin{equation}
\label{*7}
f_{\zeta}(\zeta(x),w) =
f_{\zeta}(\Phi_{*},w) \mid_{w=0}
\exp ( \Gamma \frac{ ( \zeta - \Phi_{*} ) }
{ \Phi_{*} } )
 \exp( w \lambda ) \ \ ,
\end{equation}
where
\begin{equation} \label{*8}
\Gamma (w)  = -\Phi_{*}
\frac{d \Delta F(\zeta , w )}{d \zeta }  \mid_{\zeta=\Phi_{*}} \ \
.
\end{equation}
The dependence of $\Gamma$ on $w$ is rather  weak  and one can put
\begin{equation}
\Gamma(w) = \Gamma \mid_{w=0}
\end{equation}
for some essential part of  activities spectrum.
What is the term "essential part of activities spectrum" will be clear
later.

With the help of  the conservation laws for  heterogeneous centers
and for  molecules of the
substance   one can get for \( g,  \theta \)
\begin{equation}
g(z,w) = f_{*}  \int_{0}^{z} (z-x)^{3}
\exp ( -\Gamma \frac{ g^{tot}  }
{ \Phi_{*} } )
\theta (x,w)  dx \eta_{tot}(w) \exp( w \lambda )
\equiv
G_w (g^{tot}, \theta) \ \ ,
\end{equation}
\begin{equation}\label{q17}
g^{tot} = \int dw g(z,w) \ \ ,
\end{equation}
\begin{equation}\label{q18}
\theta(z,w)
 = \exp ( - f_{*}\exp(  \lambda w)  n_{\infty}  \int_{0}^{z}
\exp (  - \Gamma \frac{ g^{tot}  }
{ \Phi_{*} } ) dx )
\equiv
S_w ( g^{tot}) \ \ ,
\end{equation}
where $f_{*} = f_{\zeta}(\Phi_{*},w=0)
$.

The droplets size spectrum  can be found as
\begin{equation}
f(x,w) = f_{*} \exp(\lambda w)
\exp ( -\Gamma \frac{ \int  dw g(x,w)  }
{ \Phi_{*} } )
\theta(x,w)
\eta_{tot} (w) \ \ ,
\end{equation}
\begin{equation} \label{11}
f_{i}=
f_{*\ i}
\exp ( -\Gamma_{i} \frac{ \sum_{j}g_{j}  }
{ \Phi_{*} } )
\theta_{i} \ \ .
\end{equation}

Since  the accuracy of the theory  is measured in the terms of
the error in the droplets number $N_i$,
 these values are defined as
\begin{equation} \label{10}
N_{i} = \eta_{tot\ i} ( 1 - \theta_{i}(z)) \equiv Q_{i}(\theta_{i})
\ \ .
\end{equation}

The structure of further constructions will be the following
\begin{itemize}
\item
 At first the asymptotic cases will be investigated. It will be done for
two types of heterogeneous centers. The generalization for many types
of centers is evident and trivial.
Here it isn't assumed that $\Gamma_i$ don't depend on $i$.
\item
Then  the general
intermediate situation will be constructed
with the help of approximation $\Gamma_{i} = const$
for all sorts of centers. One can prove that
this intermediate situation and the asymptotic
cases cover all possible situations.
\item
Then the case of the quasi continuous spectrum of the heterogeneous
centers activities will be
considered. Here one can use some
more elegant approaches which will lead to more compact results.
\end{itemize}

\section{Asymptotic solutions}

\subsection{Formal generalization of iteration method}

At first one has to show that the case under consideration can not be reduced to trivial
generalization of known approaches.
The formal generalization of the iteration method analogous to \cite{LettMathPhys}
for  the condensation on  one given sort of  heterogeneous
centers
leads to the following equations:
\begin{equation} \label{12}
g_{i\ (l+1)} = G_{i}(\sum_{j}g_{j\ (l)} , \theta_{i\ (l)})     \ \ ,
\ \ \ \
\theta_{i \ (l+1)}=S_{i}(\sum_{j} g_{j\ (l)})                        \ \ ,
\ \ \ \
N_{i\ (l)} = Q_{i}(\theta_{i\ (l)})                                        \ \ ,
\end{equation}
\begin{equation} \label{15}
g_{i\ (0)} = 0\ \ ,
\ \ \ \
\theta_{i\ (0)} = 1 \ \ ,
\end{equation}
\begin{equation} \label{17}
g_{i\ (1)} = f_{*\ i} \frac{z^{4}}{4}                                           \ \ ,
\ \ \ \
\theta_{i\ (1)} = \exp(-f_{*\ i} \frac{n_{\infty}}{\eta_{tot\ i}}
 z)           \ \ ,
\end{equation}
\begin{equation} \label{19}
N_{i\ (2)}(\infty) = \eta_{tot \ i}
[ 1 -\exp(-f_{*\ i} \frac{n_{\infty}}{\eta_{tot\ i}}
(\sum_{j} \frac{\Gamma_{i}f_{*\ j}}{4\Phi_{*}})^{-1/4} A)] \ \ ,
\ \ \ \
A = \int_{0}^{\infty} \exp(-x^{4}) dx \approx 0.9                 \ \ .
\end{equation}
The third iteration can not be calculated in the analytical form.

Let us analyze  \( N_{i(2)}({\infty}) \). Assume
 that $\zeta$ is fixed and for some $i$ and $j$
$$ f_{*\ i}   \gg f_{*\ j} \ \ .  $$
Let us decrease \( \eta_{tot\ i} \) and $\Delta_i F$
keeping the constant value of $ f_{*\ i} \sim \eta_i \exp(-\Delta_i F) $.
It is obvious that when $ \eta_{tot\ i} $  is small
then the total number of
the heterogeneously
formed droplets
coincides with
the total quantity of
 heterogeneous centers and goes to zero when $\eta_{tot\ i}$ goes to zero.
The value of $ g_{i}$ at
the end of the period of the droplets formation
on the heterogeneous centers of the sort "$j$" can be estimated as
$$
g_{i} \leq \frac{\eta_{tot\ i} (\hat{\Delta} x_{j})^{3}}{n_{\infty}}
\ \ ,
$$
where $ \hat{\Delta} x_{j} $ is the width of the size spectrum (of the size distribution
function) for  the droplets of the  sort "$j$".
The value of $ \hat{\Delta} x_{j}$ is  restricted  from above by the
value $ \Delta x_{j} $ which is the width of the size
spectrum  without any  influence of
the droplets of the other sorts and without any
exhaustion of the heterogeneous centers of
this  sort.
 Then the influence
of the heterogeneous centers of the sort $i$ on the process of the condensation
on the centers of the sort $j$ becomes
negligible in the limit $ \eta_{tot\ i} \rightarrow 0 $.
At the same time (\ref{19}) shows that in the
limit
$ \eta_{tot\ i} \rightarrow 0 $, $ f_{*\ i}  = const $  the influence of
the  droplets of the sort $i$
doesn't become small. This leads to
the big  error for $N_{j}(\infty)$ in the second  approximation.

One can not get an analytical expression in the third
approximation for $N_i$ in   frames of the standard iteration method and
the second iteration gives the wrong qualitative results. This is the
main disadvantage of the standard iteration procedure.
The reason for  this disadvantage
is the following
one. Consider the situation with one type of centers. In the case when the interruption
 of the embryos formation is caused by
exhaustion of  heterogeneous centers the error in the value of
$g_i$ is compensated by the squeezing force of the operator $S_i$.
The analogous property is absent for the operator $Q_i$ in the
situation with several types of centers due to the cross influence
of the droplets formed on different sorts. This shows that the
situation with several types of centers can not be effectively
described by a formal generalization of the already known methods.

\subsection{Characteristic lengths}

The direct  generalization of the iteration method fails
due to the wrong account of the cross
influence of the  droplets formed on  different sorts.
Nevertheless it allows
to get the spectrum of the droplets
 when the cross influence is excluded.

On a base of the first iterations in the  general procedure one can see that
 for a separate process
there are two characteristic lengths. The first one is the
length of spectrum in the situation when
there are no exhaustion of the heterogeneous
centers (and no droplets of the other sort).
One can say that the condensation occurs in the pseudo-homogeneous
way. For this characteristic value
\begin{equation} \label{22}
\Delta_{i}x = (\frac{4\Phi_{*}}{\Gamma_{i} f_{*\ i}})^{1/4}
\ \ .
\end{equation}
This length is going from the first iteration for $g_{i}$.
The second length is
the length of the spectrum
the spectrum is formed
only by
 exhaustion of  heterogeneous centers. Then the width of
the spectrum is
\begin{equation} \label{23}
\delta_{i}x = \frac{\eta_{tot\ i}}{f_{*\ i} n_{\infty} }
\ \ .
\end{equation}
This length is going from the first iteration for $\theta_{i}$.

Practically the hierarchy between $\Delta_{i}x$ and $ \delta_{i}x$ is ensured by
the hierarchy between
$f_{*\ i}$ and $\eta_{*\ i}$.  The values of $\Gamma_{i}$ are
rather (in comparison with $f_{*\ i}$) unsensible to the value of the
supersaturation. Really
\begin{equation} \label{24}
-\frac {\Gamma_{i}}{\zeta} = \frac{d\Delta_i F}{d \zeta} \sim
 \frac{d F_{i\ c}}{d \zeta}-  \frac{d F_{i\ e}}{d \zeta}
 \ \ .
\end{equation}
The value $dF_{i\ c}/d\zeta$
can be  estimated from above
by  the value in the limit of  homogeneous condensation
  $dF_{c\ hom} / d\zeta$
when the force of interaction between the heterogeneous center and molecules
of liquid decreases monotonously in space.
Since the energy of the solvatation\footnote{Under
barrier character of condensation, i.e. when $\Delta_i F \gg 1$.}
depends on the supersaturation
weaker than  $F_{c}$ depends, one can
neglect the last term of the previous equation and get
\begin{equation} \label{25}
 \frac{d\Delta F}{d \zeta} \sim
 \frac{d F_{c\ hom}}{d \zeta}
 \ \ .
\end{equation}
This dependence is  rather weak  in comparison with a very sharp
dependence of
$f_{*\ i}$  on  the supersaturation.

Another  important fact
 is the frontal form of a back side of  spectrum\footnote{The front side has
evidently the front character.} in the
pseudo-homogeneous situation (when $\Gamma_{i} $ really plays an important
role)\footnote{The
pseudo homogeneous situation can be defined  as the situation when the centers
of condensation remain practically free.}.
The frontal character can be seen from
\begin{equation} \label{26}
f_{i} = f_{*\ i} \exp(- \frac{\Gamma_{i}}{4 \Phi_{*}}(\sum_{j}f_{*\ j}) z^{4})
\ \ .
\end{equation}
A moderate variation of  $\Delta_{i}x$
can be caused only by a very big variation
 of $f_{*\ i}$.

Instead of $\delta_{i} x $ one can use  parameter
\begin{equation} \label{27}
h_{i} = \frac{\delta_{i}x}{\Delta_{i}x}
\end{equation}
to simplify formulas.

Now one can directly analyze all asymptotic cases.

\subsection{The case $\Delta_{1}x \sim \Delta _{2}x$}

\subsubsection{Situation $h_{1} \ll 1,\ \ h_{2} \geq 1$}

In this situation
\begin{itemize}
\item
The process of  formation of the droplets on  heterogeneous centers
of the first sort  doesn't depend
on  formation of the droplets of the second sort.
\end{itemize}
It can be
directly seen from the chain of inequalities
\begin{equation} \label{29}
\delta_{1} x \ll \Delta_{1} x \sim \Delta_{2} x \leq \delta_{2} x
\ \ .
\end{equation}
So, the process of  formation of the first sort droplets is described
 by the
following equalities:
\begin{equation} \label{30}
g_{1}=f_{*\ 1}
\int_{0}^{z} (z-x)^{3} \exp(-\Gamma_{1} \frac{g_{1}(x)}{\Phi_{*}}) \theta_{1}
dx
\equiv
G_{1}(g_{1},\theta_{1})
\ \ ,
\end{equation}
\begin{equation} \label{31}
\theta_{1} = \exp(- f_{*\ 1} \frac{n_{\infty}}{\eta_{tot\ 1}}
\int_{0}^{z} \exp(-\Gamma_{1} \frac{g_{1}(x)}{\Phi_{*}})
dx )
\equiv
S_{1}(g_{1})
\ \ .
\end{equation}
The system (\ref{30})  - (\ref{31})   can be reduced
by rescaling to
\begin{equation} \label{30*}
g_{1}=
\int_{0}^{z} (z-x)^{3} \exp(-g_{1}(x)) \theta_{1}
dx
\ \ ,
\end{equation}
\begin{equation}\label{31*}
\theta_{1} = \exp(- a
\int_{0}^{z} \exp(-\Gamma_{1} \frac{g_{1}(x)}{\Phi_{*}})
dx )
\end{equation}
with some parameter $a= f_*^{3/4} n_{\infty} \Phi_*^{1/4} / (\eta_{tot\ 1} \Gamma_1^{1/4})$.
 Solution of the system  (\ref{30*})-(\ref{31*})
 is drawn in Fig. 1  for different $a$.
\begin{figure}[hgh]


{\small \it
\begin{center}
Figure 1.
\\
Solution of the system (\ref{30*}) - (\ref{31*}).
The  behavior  of $f = \exp(-g) \theta $ as function of $x$
   is drawn.
   \end{center}
}
\end{figure}

 The system (\ref{30}) - (\ref{31})  can be solved by the following iterations
\begin{equation} \label{32}
g_{1\ (i+1)} = G_{1}(g_{1\ (i)},\theta_{1\ (i)})           \ \ ,
\ \ \ \
\theta_{1\ (i+1)}=S_{1}(g_{1\ (i)})                              \ \ ,
\ \ \ \
N_{1\ (i)} = Q_{1}(\theta_{1\ (i)})                                    \ \ .
\end{equation}
When for all values of the arguments
$$ w_{1} \leq w_{2}$$
then
$$ S_{1}(w_{1}) \leq S_{1}(w_{2}) \ \ .$$
When  for all values of the arguments
$$ w_{1} \leq w_{2}$$
then
$$ Q_{1}(w_{1}) \geq Q_{1}(w_{2}) \ \ \ .$$
When  for all values of the arguments
$$ w_{1} \leq w_{2}
\ \ , \ \ \ \  v_{1} \geq v_{2}$$
then
$$ G_{1}(v_{1},w_{1}) \leq G_{1}(v_{2},w_{2})  \ \ .$$

The initial approximations can be chosen as
\begin{equation} \label{35}
g_{1\ (0)} = 0 \ ,
\ \ \ \
\theta_{1\ (0)} = 1 \ \ ,
\end{equation}
which leads to
\begin{equation}  \label{chain}
 g_{1\ (0)} \leq g_{1\ (2)} \leq  ... \leq  g_{1\ (2i)}\leq ... \leq g_{1} \leq
... \leq
 g_{1\ (2i+1)} \leq ... \leq g_{1\ (3)} \leq g_{1\ (1)} \ ,
 \end{equation}
$$ \eta_{1\ (0)} \geq \eta_{1\ (2)} \geq  ... \geq
\eta_{1\ (2i)} \geq ... \geq \eta_{1} \geq
... \geq
 \eta_{1\ (2i+1)} \geq ... \geq \eta_{1\ (3)} \geq \eta_{1\ (1)} \ ,  $$
\begin{equation}  \label{chain1}
 N_{1\ (0)} \leq N_{1\ (2)} \leq  ... \leq  N_{1\ (2i)}\leq ... \leq N_{1} \leq
... \leq
 N_{1\ (2i+1)} \leq ... \leq N_{1\ (3)} \leq N_{1\ (1)} \ .
 \end{equation}
These estimates allow to prove the convergence of  iterations.

The calculation of  iterations gives
\begin{equation} \label{38}
g_{1\ (1)} = f_{*\ 1} \frac{z^{4}}{4} \ ,
\ \ \  \
\theta_{1\ (1)} = \exp(-f_{*\ 1}
 \frac{n_{\infty}}{\eta_{tot\ 1}}z) \ ,
\end{equation}
\begin{equation} \label{40}
N_{1\ (2)}(\infty) = \eta_{tot\ 1} ( 1-\exp(-f_{*\ 1}
\frac{n_{\infty}}{\eta_{tot\ 1}}
 ( \frac{\Gamma_{1}}
{4 \Phi_{*}})^{-1/4}  f_{*\ 1}^{-1/4} A ))   \ .
\end{equation}
Since
\begin{equation} \label{41}
\frac{d}{dx} \mid N_{1\ (i)}-N_{1\ (j)} \mid  \geq 0 \ \ ,
\end{equation}
 by
the simple numerical calculation of $ N_{1(3)}(\infty)$  one can
see
  $$\frac{ \mid N_{1\ (2)} - N_{1} \mid }{N_{1}} \leq 0.015 \ , $$
which means that the second iteration is already rather precise\footnote{
The same procedure can be used for the condensation on one sort of heterogeneous
centers.}.

On the base of the iterations one can get some approximations for the
supersaturation\footnote{Without the second sort droplets taken into account.}:
\begin{equation} \label{42}
\zeta_{(l+1)} = \Phi_{*} - f_{*\ 1} \int_{0}^{z} (z-x)^{3}
\exp(-\Gamma_{1}\frac{g_{1\ (l)}}{\Phi_{*}}) \theta_{1\ (l)} dx
\end{equation}
 and  come to
the second approximation for $\zeta$
\begin{equation} \label{43}
\zeta_{(2)} = \Phi_{*} - f_{*\ 1} \int_{0}^{z} (z-x)^{3}
\exp(-H x) dx
 =
\Phi_* + f_{*\ 1} [ - \frac{z^3}{H} + \frac{3z^2}{H^2} - \frac{6 z}{H^3} + \frac{6}{H^4} \exp(-Hz)]
\ \ ,
\end{equation}
where
\begin{equation} \label{44}
H = \frac{f_{*\ 1} n_{\infty}}{\eta_{tot\ 1}}
\end{equation}

This expression can be simplified. The value of supersaturation
appears in expression for the size spectrum $f(x)$ in the
 form
$ \exp(-\Gamma_{i} (\zeta-\Phi_{*})/\Phi_{*})$ .
After the substitution of $\zeta_{(2)}$ into this expression one can
see that in the case when $\zeta$  deviates  essentially (i.e. when the
exponent changes) from $\Phi_*$
all  terms except the first two ones can be neglected and
\begin{equation} \label{46}
\zeta_{(2)} = \Phi_{*} - z^3 \frac{\eta_{tot\ 1}}{n_{\infty}} \ \ .
\end{equation}

For the second sort one can obtain the following system of equations
\begin{equation} \label{47}
g_{2}=f_{*\ 2}
\int_{0}^{z} (z-x)^{3} \exp(-\Gamma_{2} \frac{g_{2}(x)
+(\eta_{tot\ 1}/n_{\infty})x^3
}{\Phi_{*}}) \theta_{2} dx
\equiv
G_{2}(g_{2}
+(\eta_{tot\ 1}/n_{\infty})x^3,\theta_{2}) \ \ ,
\end{equation}
\begin{equation} \label{48}
\theta_{2} = \exp(- f_{*\ 2} \frac{n_{\infty}}{\eta_{tot\ 2}}
\int_{0}^{z} \exp(-\Gamma_{2} \frac{g_{2}(x)
+(\eta_{tot\ 1}/n_{\infty})x^3
}{\Phi_{*}})
dx )
\equiv
S_{2}(g_{2}
+(\eta_{tot\ 1}/n_{\infty})x^3) \ \ .
\end{equation}
This system can be rescaled to
\begin{equation} \label{47*}
g_{2}=
\int_{0}^{z} (z-x)^{3} \exp(-g_{2}(x)
- b x^3) \theta_{2} dx \ \ ,
\end{equation}
\begin{equation} \label{48*}
\theta_{2} = \exp(- a
\int_{0}^{z} \exp(-g_{2}(x)
- b x^3)
dx )
\end{equation}
with constants $a= f_{*\ 2}^{3/4} n_{\infty} \Phi_*^{1/4} / (\eta_{tot\ 2} \Gamma_2^{1/4})$
and $b = \eta_{tot\ 2} \Gamma_2^{3/4} /( n_{\infty} \Phi_*^{3/4} f_{*\ 2}^{1/4})$.
Solutions of this system for different values of $a$ and $b$ are
shown in Fig. 2.
\begin{figure}



{\small \it
\begin{center}
Figure 2.
\\
Solution of the system (\ref{47*}) - (\ref{48*}).
The  behavior  of $f = \exp(-g) \theta $ as function of $x$
   is drawn.
   \end{center}
}

\end{figure}

Having introduced
\begin{equation} \label{49}
\lambda_{2}=g_{2}
+(\eta_{tot\ 1}/n_{\infty})z^3 \ \ ,
\end{equation}
one can rewrite the system (\ref{47})-(\ref{48}) as
\begin{equation} \label{50}
\lambda_{2} = G_{2}(\lambda_{2},\theta_{2})
+(\eta_{tot\ 1}/n_{\infty})z^3
\equiv
G_{2}^{+}(\lambda_{2},\theta_{2}) \ \ ,
\end{equation}
\begin{equation} \label{51}
\theta_{2} = S_{2}(\lambda_{2}) \ \ .
\end{equation}

The operator $G^{+}_{2}$ has the same properties as $G_{1},G_{2}$ have.
All chains of inequalities
remain valid with the index "$2$"
instead of "$1$" and the operator $G^{+}_2$ instead of $G_2$.
Moreover one can see that
\begin{equation} \label{52}
\frac{d}{d(\eta_{tot\ 1}/n_{\infty})} \mid N_{2\ (i)} - N_{2\ (j)} \mid \leq 0
\ \ ,
\end{equation}
which shows that the worst situation for the iterations convergence is
$\eta_{tot} = 0$. Even in this situation the second iteration is
rather precise. It can be seen from    investigation of nucleation
on the first sort
of centers.

Actually  one can avoid here the
calculations according to  such a complex procedure.
The term $\eta_{tot\ 1} z^3/ n_{\infty}$ ensures
the characteristic length
\begin{equation} \label{53}
D_{1}=(\frac{\Phi_{*} n_{\infty}}{\Gamma_{2} \eta_{tot\ 1}})^{1/3}
\ .
\end{equation}
Since
\begin{equation} \label{54}
D_{1} \geq \epsilon \Delta_{1} x \sim \epsilon \Delta_{2} x \ \ \ \ \
\epsilon\sim (2 \div 3 ) \ \ ,
\end{equation}
the condensation on the centers of the second sort occurs in the separate
way,
the influence of the first sort of centers here
is negligible and one can
use the  formulas (\ref{30})-(\ref{32}), (\ref{38})-(\ref{41})
 with the
index "$2$" instead of the index "$1$".

\subsubsection{Situation $h_{1} \geq 1,\ \ \ h_{2}\ll 1$}

Since  $\Delta_{1} x  \sim \Delta_{2}  x $, one can change the numbers of sorts and
reduce this situation to the previous one.

\subsubsection{Situation $h_{1} \geq 1,\ \ \ h_{2} \geq 1$}

To analyze this situation one has to understand why in the separate
condensation on one sort already the second iteration gives
rather precise results.
This property  is explained by the big power $3$ in
the subintegral expression for $g_i$. Then the droplets
of the big sizes
near the front side of the spectrum are the main consumers of the
vapor. These droplets have
small (in comparison with $\Delta_{i}x$) values of the variable $x$.
 The cross influence is
rather weak and one can use the general iteration
procedure which gives
\begin{equation} \label{55}
g_{i\ (1)} = f_{*\ i} \frac{z^{4}}{4} \ \ , \ \ i=1,2\ ,
\end{equation}
\begin{equation} \label{56}
\theta_{i\ (1)} = \exp(-f_{*\ i} \frac{n_{\infty}}{\eta_{tot\ 1}}
 z) \ \ ,
\end{equation}
\begin{equation} \label{57}
N_{i\ (2)}(\infty) = \eta_{tot \ i}
[ 1 -\exp(-f_{*\ i} \frac{n_{\infty}}{\eta_{tot\ i}}
( (\frac{\Gamma_{i}}{4\Phi_{*}})f_{*\ 1}+
(\frac{\Gamma_{i}}{4\Phi_{*}})f_{*\ 2})^{-1/4}A)] \ \ .
\end{equation}
Since the exhaustion of heterogeneous centers is moderate, the
precision of iterations resembles the homogeneous case and
allows the estimate $|N_{i\ (2)} - N_i | < 0.15 N_i$.

\subsubsection{Situation $h_{1}\ll 1, h_{2} \ll 1$}

Since
\begin{equation} \label{61}
\delta_{1}x \ll \Delta_{1}x \sim \Delta_{2}x \ \ ,
\end{equation}
 the droplets  of the
second sort don't act on the process of formation  of
the droplets of the first sort.
Since
\begin{equation} \label{62}
\delta_{2} x \ll \Delta_{2}x \sim \Delta_{1}x \ \ ,
\end{equation}
the same  is
valid for the droplets of  the second sort.
The system is split into parts corresponding to the
separate processes of the condensation on  different sorts.
Eq. (\ref{30})-(\ref{32}) can be reproduced here.
But in the case $h_{i} \ll 1$
for all $i$ the vapor exhaustion can be neglected in comparison
with the exhaustion of  heterogeneous centers
and one can get
precise explicit results
\begin{equation}\label{63}
\theta_{i}(x) = \exp(-f_{*\ 1} \frac{n_{\infty}}{\eta_{tot\ 1}} z)
\ \ , \ i=1,2\ \ ,
\end{equation}
\begin{equation}\label{65}
f_{i}(x) = f_{*\ i} \exp(-f_{*\ i} \frac{n_{\infty}}{\eta_{tot\ i}} z)
\ \ ,
\end{equation}
\begin{equation} \label{67}
\zeta = \Phi_{*} - (\frac{\eta_{tot\ 1}}{n_{\infty}} +
\frac{\eta_{tot\ 2}}{n_{\infty}}) z^3 \ \ .
\end{equation}
This expression for $\zeta$ is obtained by the same procedure
as that which led to (\ref{46}).

\subsection{Case $\Delta_{1} x \ll \Delta_{2} x$}

Due to $\Delta_{1} x \ll \Delta_{2} x$
the droplets of the second sort don't
act on the process of formation of the droplets of the first sort.
The process\footnote{The case
$\Delta_2 x  \gg \Delta_1 x $
is reversal to this case and can be considered by the simple change
of the indexes.
} of
 formation of the droplets of the first sort can be described by
the iteration procedure  presented by eq.(\ref{32}).

\subsubsection{Situation $h_{1}\ll 1, h_{2} \geq 1 $}

Due to $h_{1}\ll 1$ the equation for $g_1$ can be
simplified
\begin{equation} \label{68}
g_{1}= f_{*\ 1} \int_{0}^{z} (z-x)^3 \exp(-Hx) dx \sim
\frac{\eta_{tot \ 1}}{n_{\infty}} z^3 \ \ .
\end{equation}
The value of $\theta_{1}$ is given by (\ref{63}),
 the value of $f_{1} (x) $ is given by (\ref{65}).

For the condensation on the centers of the
second sort  the  equations analogous to (\ref{47}),(\ref{48}) are suitable.
So, one can get here equations (\ref{49})-(\ref{52}).
But in this situation the inequality (\ref{54})
isn't valid and one must calculate  iterations.
One can choose as initial approximation $\lambda_{2\ (0)} = 0$ and get
\begin{equation} \label{69}
\lambda_{2\ (1)} = f_{*\ 2}\frac{z^{4}}{4} +
\frac{\eta_{tot\ 1}}{n_{\infty}} z^{3} \ \ ,
\end{equation}
\begin{equation} \label{70}
\theta_{2\ (2)}=
\exp(- f_{*\ 2}
\frac{n_{\infty}}{\eta_{tot\ 2}} \int_{0}^{z}
\exp(-( \frac{x}{\Delta_{\infty\ 2}x})^{4} - (\frac{x}{\Delta_{h\ 1} x })^{3})
dx) \ \ ,
\end{equation}
where
$$ \Delta_{\infty \ 2} x = (\frac{4 \Phi_{*}}{\Gamma_{2} f_{*\ 2}})^{1/4}
\equiv
\Delta_2 x
\ \ , \ \ \ \ \  \Delta_{h \ 1} x = ( \frac{\Phi_{*}n_{\infty}}
{\Gamma_{2} \eta_{tot\ 1}} )^{1/3}
\equiv
D_1 \ \ .
$$
Eq. (\ref{70}) can be rescaled to
\begin{equation} \label{70*}
\theta_{2\ (2)}=
\exp(- a  \int_{0}^{z}
\exp(-x^{4} - b x^{3})
dx)
\end{equation}
with parameters $a = f_{*\ 2} n_{\infty} \Delta_{\infty \ 2} x / \eta_{tot\ 2}$ and
$b= (\Delta_{\infty\ 2} x )^3 / (\Delta_{1\ h} x)^3$.
This dependence is shown in Fig. 3.

\begin{figure}



{\small \it
\begin{center}
Figure 3.
\\
Behavior  of $\theta$ as function of $x$
for different $a$, $b$.
   \end{center}
}

\end{figure}

The value of $\Delta_{\infty \ 2} x$ has
the sense of the spectrum width when the
cross influence and the  exhaustion  of  heterogeneous
centers
are neglected. The value of $\Delta_{h \ 1} x $
 has the sense of the spectrum width
when the vapor consumption  by  droplets is neglected.

One can easily prove that
$$ \frac{d}{dx} \mid N_{2\ (i)} - N_{2\ (j)} \mid \geq 0
\ \ , \ \ \ \
 \frac{d}{d\eta_{tot\ 1}} \mid N_{2\ (i)} - N_{2\ (j)} \mid \leq 0 $$
for $i,j \geq 2$.
Then it is easy to show that
$$ \frac{\mid N_{2\ (2)} - N_{2} \mid}{N_{2}} \leq 0.015 $$
by the calculation of  $N_{2\ (2)}(\infty)$ and
$N_{2\ (3)}(\infty)$ at $\eta_{tot\ 1} = 0 $.

The simple approximation  can be
obtained
if one notices that on the base of (\ref{70})
\begin{equation} \label{71}
\theta_{2\ (2)}(\infty)
\approx
 \exp[-f_{*\ 2} \Delta_{\infty \ 2} x
\frac{n_{\infty}}{\eta_{tot\ 2}}
(\frac{A}{2}(1+ (\frac{\Delta_{\infty \ 2}x}{\Delta_{h \ 1}x})^{4})^{-1/4} +
\frac{B}{2}(1+ (\frac{\Delta_{\infty \ 2 }x}{\Delta_{h\ 1}x})^{3})^{-1/3})]
\ \ ,
\end{equation}
where
$$ B = \int_{0}^{\infty}\exp(-x^{3})
dy \ \ , $$
with the relative error less than 0.035.

The spectrum of sizes of the droplets formed on the centers
of the second sort is
\begin{equation} \label{72}
f_{2} = f_{*\ 2} \exp(-\frac{\Gamma_{2} f_{*\ 2}}{\Phi_{*}} \frac{z^4}{4})
\exp(-\frac{\Gamma_2}{\Phi_*} \frac{\eta_{tot 1}}{n_{\infty}} z^3)
\exp( - f_{*\ 2}\frac{n_{\infty}}{\eta_{tot\ 2}} \int_{0}^{z}
\exp( - (\frac{x}{\Delta_{\infty \ 2 }x})^{4} - (\frac{x}{\Delta_{h \ 1} x
})^{3}) dx) \ \ .
\end{equation}
By appropriate rescaling it can be reduced to
\begin{equation} \label{72*}
f_{2} \sim \exp(- z^4)
\exp(-(\frac{ z}{b})^3)
\exp( - a \int_{0}^{z}
\exp( - x^{4} - (\frac{x}{ b
})^{3}) dx)
\end{equation}
with two parameters $
a =
f_{*\ 2} n_{\infty} \Delta_{\infty \ 2} x / \eta_{tot\ 2} $
and $
b=
\Delta_{h \ 1} x / \Delta_{\infty \ 2} x
$ which is drawn in Fig.4.

\begin{figure}

\begin{picture}(450,250)
\put(210,170){\vector(-1,-1){91}}
\put(227,145){\vector(-1,-1){84}}
\put(177,175){\vector(-1,-1){59}}
\put(215,195){\vector(-1,-1){83}}
\put(210,173){$a=2,b=0.5$}
\put(190,199){$a=0.5,b=2$}
\put(213,149){$a=2,b=2$}
\put(107,177){$a=0.5,b=0.5$}
\put(60,10){\special{em: moveto}}
\put(290,10){\special{em: lineto}}
\put(290,240){\special{em: lineto}}
\put(60,240){\special{em: lineto}}
\put(60,10){\special{em: lineto}}
\put(65,15){\special{em: moveto}}
\put(285,15){\special{em: lineto}}
\put(285,235){\special{em: lineto}}
\put(65,235){\special{em: lineto}}
\put(65,15){\special{em: lineto}}
\put(100,50){\vector(1,0){150}}
\put(100,50){\vector(0,1){150}}
\put(150,50){\vector(0,1){3}}
\put(200,50){\vector(0,1){3}}
\put(100,100){\vector(1,0){3}}
\put(100,150){\vector(1,0){3}}
\put(90,40){$0$}
\put(148,40){$1$}
\put(198,40){$2$}
\put(82,98){$0.5$}
\put(90,148){$1$}
\put(108,208){$f$}
\put(255,55){$x$}
\put(100,150){\special{em: moveto}}
\put(100.50,149.50){\special{em: lineto}}
\put(101.00,149.00){\special{em: lineto}}
\put(101.50,148.49){\special{em: lineto}}
\put(102.00,147.97){\special{em: lineto}}
\put(102.50,147.43){\special{em: lineto}}
\put(103.00,146.88){\special{em: lineto}}
\put(103.50,146.30){\special{em: lineto}}
\put(104.00,145.69){\special{em: lineto}}
\put(104.50,145.05){\special{em: lineto}}
\put(105.00,144.37){\special{em: lineto}}
\put(105.50,143.65){\special{em: lineto}}
\put(106.00,142.89){\special{em: lineto}}
\put(106.50,142.08){\special{em: lineto}}
\put(107.00,141.22){\special{em: lineto}}
\put(107.50,140.31){\special{em: lineto}}
\put(108.00,139.34){\special{em: lineto}}
\put(108.50,138.32){\special{em: lineto}}
\put(109.00,137.24){\special{em: lineto}}
\put(109.50,136.09){\special{em: lineto}}
\put(110.00,134.89){\special{em: lineto}}
\put(110.50,133.62){\special{em: lineto}}
\put(111.00,132.29){\special{em: lineto}}
\put(111.50,130.89){\special{em: lineto}}
\put(112.00,129.43){\special{em: lineto}}
\put(112.50,127.90){\special{em: lineto}}
\put(113.00,126.31){\special{em: lineto}}
\put(113.50,124.66){\special{em: lineto}}
\put(114.00,122.95){\special{em: lineto}}
\put(114.50,121.19){\special{em: lineto}}
\put(115.00,119.37){\special{em: lineto}}
\put(115.50,117.49){\special{em: lineto}}
\put(116.00,115.57){\special{em: lineto}}
\put(116.50,113.61){\special{em: lineto}}
\put(117.00,111.60){\special{em: lineto}}
\put(117.50,109.56){\special{em: lineto}}
\put(118.00,107.49){\special{em: lineto}}
\put(118.50,105.39){\special{em: lineto}}
\put(119.00,103.27){\special{em: lineto}}
\put(119.50,101.13){\special{em: lineto}}
\put(120.00,98.99){\special{em: lineto}}
\put(120.50,96.84){\special{em: lineto}}
\put(121.00,94.70){\special{em: lineto}}
\put(121.50,92.56){\special{em: lineto}}
\put(122.00,90.44){\special{em: lineto}}
\put(122.50,88.34){\special{em: lineto}}
\put(123.00,86.26){\special{em: lineto}}
\put(123.50,84.22){\special{em: lineto}}
\put(124.00,82.21){\special{em: lineto}}
\put(124.50,80.25){\special{em: lineto}}
\put(125.00,78.34){\special{em: lineto}}
\put(125.50,76.47){\special{em: lineto}}
\put(126.00,74.67){\special{em: lineto}}
\put(126.50,72.92){\special{em: lineto}}
\put(127.00,71.24){\special{em: lineto}}
\put(127.50,69.63){\special{em: lineto}}
\put(128.00,68.09){\special{em: lineto}}
\put(128.50,66.62){\special{em: lineto}}
\put(129.00,65.22){\special{em: lineto}}
\put(129.50,63.89){\special{em: lineto}}
\put(130.00,62.65){\special{em: lineto}}
\put(130.50,61.47){\special{em: lineto}}
\put(131.00,60.37){\special{em: lineto}}
\put(131.50,59.35){\special{em: lineto}}
\put(132.00,58.39){\special{em: lineto}}
\put(132.50,57.51){\special{em: lineto}}
\put(133.00,56.70){\special{em: lineto}}
\put(133.50,55.95){\special{em: lineto}}
\put(134.00,55.27){\special{em: lineto}}
\put(134.50,54.65){\special{em: lineto}}
\put(135.00,54.08){\special{em: lineto}}
\put(135.50,53.57){\special{em: lineto}}
\put(136.00,53.11){\special{em: lineto}}
\put(136.50,52.70){\special{em: lineto}}
\put(137.00,52.33){\special{em: lineto}}
\put(137.50,52.01){\special{em: lineto}}
\put(138.00,51.72){\special{em: lineto}}
\put(138.50,51.47){\special{em: lineto}}
\put(139.00,51.25){\special{em: lineto}}
\put(139.50,51.06){\special{em: lineto}}
\put(140.00,50.89){\special{em: lineto}}
\put(140.50,50.75){\special{em: lineto}}
\put(141.00,50.62){\special{em: lineto}}
\put(100,150){\special{em: moveto}}
\put(100.50,149.50){\special{em: lineto}}
\put(101.00,149.00){\special{em: lineto}}
\put(101.50,148.51){\special{em: lineto}}
\put(102.00,148.02){\special{em: lineto}}
\put(102.50,147.53){\special{em: lineto}}
\put(103.00,147.04){\special{em: lineto}}
\put(103.50,146.55){\special{em: lineto}}
\put(104.00,146.07){\special{em: lineto}}
\put(104.50,145.58){\special{em: lineto}}
\put(105.00,145.10){\special{em: lineto}}
\put(105.50,144.62){\special{em: lineto}}
\put(106.00,144.14){\special{em: lineto}}
\put(106.50,143.66){\special{em: lineto}}
\put(107.00,143.17){\special{em: lineto}}
\put(107.50,142.69){\special{em: lineto}}
\put(108.00,142.21){\special{em: lineto}}
\put(108.50,141.72){\special{em: lineto}}
\put(109.00,141.23){\special{em: lineto}}
\put(109.50,140.75){\special{em: lineto}}
\put(110.00,140.25){\special{em: lineto}}
\put(110.50,139.76){\special{em: lineto}}
\put(111.00,139.26){\special{em: lineto}}
\put(111.50,138.76){\special{em: lineto}}
\put(112.00,138.26){\special{em: lineto}}
\put(112.50,137.75){\special{em: lineto}}
\put(113.00,137.24){\special{em: lineto}}
\put(113.50,136.72){\special{em: lineto}}
\put(114.00,136.19){\special{em: lineto}}
\put(114.50,135.66){\special{em: lineto}}
\put(115.00,135.12){\special{em: lineto}}
\put(115.50,134.58){\special{em: lineto}}
\put(116.00,134.03){\special{em: lineto}}
\put(116.50,133.47){\special{em: lineto}}
\put(117.00,132.90){\special{em: lineto}}
\put(117.50,132.32){\special{em: lineto}}
\put(118.00,131.73){\special{em: lineto}}
\put(118.50,131.14){\special{em: lineto}}
\put(119.00,130.53){\special{em: lineto}}
\put(119.50,129.91){\special{em: lineto}}
\put(120.00,129.29){\special{em: lineto}}
\put(120.50,128.65){\special{em: lineto}}
\put(121.00,128.00){\special{em: lineto}}
\put(121.50,127.33){\special{em: lineto}}
\put(122.00,126.66){\special{em: lineto}}
\put(122.50,125.97){\special{em: lineto}}
\put(123.00,125.27){\special{em: lineto}}
\put(123.50,124.56){\special{em: lineto}}
\put(124.00,123.83){\special{em: lineto}}
\put(124.50,123.09){\special{em: lineto}}
\put(125.00,122.33){\special{em: lineto}}
\put(125.50,121.56){\special{em: lineto}}
\put(126.00,120.78){\special{em: lineto}}
\put(126.50,119.98){\special{em: lineto}}
\put(127.00,119.16){\special{em: lineto}}
\put(127.50,118.33){\special{em: lineto}}
\put(128.00,117.49){\special{em: lineto}}
\put(128.50,116.63){\special{em: lineto}}
\put(129.00,115.76){\special{em: lineto}}
\put(129.50,114.87){\special{em: lineto}}
\put(130.00,113.96){\special{em: lineto}}
\put(130.50,113.05){\special{em: lineto}}
\put(131.00,112.11){\special{em: lineto}}
\put(131.50,111.17){\special{em: lineto}}
\put(132.00,110.20){\special{em: lineto}}
\put(132.50,109.23){\special{em: lineto}}
\put(133.00,108.24){\special{em: lineto}}
\put(133.50,107.24){\special{em: lineto}}
\put(134.00,106.22){\special{em: lineto}}
\put(134.50,105.19){\special{em: lineto}}
\put(135.00,104.16){\special{em: lineto}}
\put(135.50,103.10){\special{em: lineto}}
\put(136.00,102.04){\special{em: lineto}}
\put(136.50,100.97){\special{em: lineto}}
\put(137.00,99.89){\special{em: lineto}}
\put(137.50,98.80){\special{em: lineto}}
\put(138.00,97.70){\special{em: lineto}}
\put(138.50,96.60){\special{em: lineto}}
\put(139.00,95.49){\special{em: lineto}}
\put(139.50,94.37){\special{em: lineto}}
\put(140.00,93.25){\special{em: lineto}}
\put(140.50,92.13){\special{em: lineto}}
\put(141.00,91.00){\special{em: lineto}}
\put(141.50,89.87){\special{em: lineto}}
\put(142.00,88.74){\special{em: lineto}}
\put(142.50,87.61){\special{em: lineto}}
\put(143.00,86.49){\special{em: lineto}}
\put(143.50,85.36){\special{em: lineto}}
\put(144.00,84.24){\special{em: lineto}}
\put(144.50,83.13){\special{em: lineto}}
\put(145.00,82.02){\special{em: lineto}}
\put(145.50,80.92){\special{em: lineto}}
\put(146.00,79.82){\special{em: lineto}}
\put(146.50,78.74){\special{em: lineto}}
\put(147.00,77.67){\special{em: lineto}}
\put(147.50,76.61){\special{em: lineto}}
\put(148.00,75.56){\special{em: lineto}}
\put(148.50,74.53){\special{em: lineto}}
\put(149.00,73.51){\special{em: lineto}}
\put(149.50,72.51){\special{em: lineto}}
\put(150.00,71.52){\special{em: lineto}}
\put(150.50,70.55){\special{em: lineto}}
\put(151.00,69.61){\special{em: lineto}}
\put(151.50,68.68){\special{em: lineto}}
\put(152.00,67.77){\special{em: lineto}}
\put(152.50,66.89){\special{em: lineto}}
\put(153.00,66.03){\special{em: lineto}}
\put(153.50,65.19){\special{em: lineto}}
\put(154.00,64.38){\special{em: lineto}}
\put(154.50,63.59){\special{em: lineto}}
\put(155.00,62.82){\special{em: lineto}}
\put(155.50,62.08){\special{em: lineto}}
\put(156.00,61.36){\special{em: lineto}}
\put(156.50,60.68){\special{em: lineto}}
\put(157.00,60.01){\special{em: lineto}}
\put(157.50,59.38){\special{em: lineto}}
\put(158.00,58.77){\special{em: lineto}}
\put(158.50,58.18){\special{em: lineto}}
\put(159.00,57.62){\special{em: lineto}}
\put(159.50,57.09){\special{em: lineto}}
\put(160.00,56.59){\special{em: lineto}}
\put(160.50,56.10){\special{em: lineto}}
\put(161.00,55.65){\special{em: lineto}}
\put(161.50,55.22){\special{em: lineto}}
\put(162.00,54.81){\special{em: lineto}}
\put(162.50,54.42){\special{em: lineto}}
\put(163.00,54.06){\special{em: lineto}}
\put(163.50,53.72){\special{em: lineto}}
\put(164.00,53.40){\special{em: lineto}}
\put(164.50,53.11){\special{em: lineto}}
\put(165.00,52.83){\special{em: lineto}}
\put(165.50,52.57){\special{em: lineto}}
\put(166.00,52.33){\special{em: lineto}}
\put(166.50,52.11){\special{em: lineto}}
\put(167.00,51.91){\special{em: lineto}}
\put(167.50,51.72){\special{em: lineto}}
\put(168.00,51.54){\special{em: lineto}}
\put(168.50,51.39){\special{em: lineto}}
\put(169.00,51.24){\special{em: lineto}}
\put(169.50,51.11){\special{em: lineto}}
\put(170.00,50.99){\special{em: lineto}}
\put(170.50,50.88){\special{em: lineto}}
\put(171.00,50.78){\special{em: lineto}}
\put(171.50,50.69){\special{em: lineto}}
\put(172.00,50.60){\special{em: lineto}}
\put(172.50,50.53){\special{em: lineto}}
\put(173.00,50.47){\special{em: lineto}}
\put(100,150){\special{em: moveto}}
\put(100.50,148.02){\special{em: lineto}}
\put(101.00,146.07){\special{em: lineto}}
\put(101.50,144.16){\special{em: lineto}}
\put(102.00,142.27){\special{em: lineto}}
\put(102.50,140.40){\special{em: lineto}}
\put(103.00,138.54){\special{em: lineto}}
\put(103.50,136.71){\special{em: lineto}}
\put(104.00,134.88){\special{em: lineto}}
\put(104.50,133.06){\special{em: lineto}}
\put(105.00,131.25){\special{em: lineto}}
\put(105.50,129.45){\special{em: lineto}}
\put(106.00,127.64){\special{em: lineto}}
\put(106.50,125.84){\special{em: lineto}}
\put(107.00,124.04){\special{em: lineto}}
\put(107.50,122.24){\special{em: lineto}}
\put(108.00,120.44){\special{em: lineto}}
\put(108.50,118.63){\special{em: lineto}}
\put(109.00,116.83){\special{em: lineto}}
\put(109.50,115.03){\special{em: lineto}}
\put(110.00,113.22){\special{em: lineto}}
\put(110.50,111.42){\special{em: lineto}}
\put(111.00,109.61){\special{em: lineto}}
\put(111.50,107.81){\special{em: lineto}}
\put(112.00,106.01){\special{em: lineto}}
\put(112.50,104.21){\special{em: lineto}}
\put(113.00,102.42){\special{em: lineto}}
\put(113.50,100.64){\special{em: lineto}}
\put(114.00,98.87){\special{em: lineto}}
\put(114.50,97.10){\special{em: lineto}}
\put(115.00,95.35){\special{em: lineto}}
\put(115.50,93.61){\special{em: lineto}}
\put(116.00,91.89){\special{em: lineto}}
\put(116.50,90.19){\special{em: lineto}}
\put(117.00,88.50){\special{em: lineto}}
\put(117.50,86.84){\special{em: lineto}}
\put(118.00,85.20){\special{em: lineto}}
\put(118.50,83.58){\special{em: lineto}}
\put(119.00,81.99){\special{em: lineto}}
\put(119.50,80.43){\special{em: lineto}}
\put(120.00,78.90){\special{em: lineto}}
\put(120.50,77.40){\special{em: lineto}}
\put(121.00,75.94){\special{em: lineto}}
\put(121.50,74.51){\special{em: lineto}}
\put(122.00,73.12){\special{em: lineto}}
\put(122.50,71.76){\special{em: lineto}}
\put(123.00,70.45){\special{em: lineto}}
\put(123.50,69.18){\special{em: lineto}}
\put(124.00,67.95){\special{em: lineto}}
\put(124.50,66.76){\special{em: lineto}}
\put(125.00,65.62){\special{em: lineto}}
\put(125.50,64.52){\special{em: lineto}}
\put(126.00,63.47){\special{em: lineto}}
\put(126.50,62.47){\special{em: lineto}}
\put(127.00,61.51){\special{em: lineto}}
\put(127.50,60.60){\special{em: lineto}}
\put(128.00,59.73){\special{em: lineto}}
\put(128.50,58.91){\special{em: lineto}}
\put(129.00,58.14){\special{em: lineto}}
\put(129.50,57.41){\special{em: lineto}}
\put(130.00,56.73){\special{em: lineto}}
\put(130.50,56.09){\special{em: lineto}}
\put(131.00,55.50){\special{em: lineto}}
\put(131.50,54.95){\special{em: lineto}}
\put(132.00,54.43){\special{em: lineto}}
\put(132.50,53.96){\special{em: lineto}}
\put(133.00,53.53){\special{em: lineto}}
\put(133.50,53.13){\special{em: lineto}}
\put(134.00,52.77){\special{em: lineto}}
\put(134.50,52.44){\special{em: lineto}}
\put(135.00,52.14){\special{em: lineto}}
\put(135.50,51.87){\special{em: lineto}}
\put(136.00,51.63){\special{em: lineto}}
\put(136.50,51.42){\special{em: lineto}}
\put(137.00,51.22){\special{em: lineto}}
\put(137.50,51.05){\special{em: lineto}}
\put(138.00,50.90){\special{em: lineto}}
\put(138.50,50.77){\special{em: lineto}}
\put(139.00,50.65){\special{em: lineto}}
\put(139.50,50.55){\special{em: lineto}}
\put(140.00,50.47){\special{em: lineto}}
\put(140.50,50.39){\special{em: lineto}}
\put(141.00,50.33){\special{em: lineto}}
\put(100,150){\special{em: moveto}}
\put(100.50,148.02){\special{em: lineto}}
\put(101.00,146.08){\special{em: lineto}}
\put(101.50,144.18){\special{em: lineto}}
\put(102.00,142.31){\special{em: lineto}}
\put(102.50,140.48){\special{em: lineto}}
\put(103.00,138.69){\special{em: lineto}}
\put(103.50,136.93){\special{em: lineto}}
\put(104.00,135.21){\special{em: lineto}}
\put(104.50,133.51){\special{em: lineto}}
\put(105.00,131.86){\special{em: lineto}}
\put(105.50,130.23){\special{em: lineto}}
\put(106.00,128.63){\special{em: lineto}}
\put(106.50,127.06){\special{em: lineto}}
\put(107.00,125.53){\special{em: lineto}}
\put(107.50,124.02){\special{em: lineto}}
\put(108.00,122.54){\special{em: lineto}}
\put(108.50,121.08){\special{em: lineto}}
\put(109.00,119.65){\special{em: lineto}}
\put(109.50,118.25){\special{em: lineto}}
\put(110.00,116.87){\special{em: lineto}}
\put(110.50,115.52){\special{em: lineto}}
\put(111.00,114.19){\special{em: lineto}}
\put(111.50,112.89){\special{em: lineto}}
\put(112.00,111.60){\special{em: lineto}}
\put(112.50,110.34){\special{em: lineto}}
\put(113.00,109.10){\special{em: lineto}}
\put(113.50,107.88){\special{em: lineto}}
\put(114.00,106.68){\special{em: lineto}}
\put(114.50,105.50){\special{em: lineto}}
\put(115.00,104.34){\special{em: lineto}}
\put(115.50,103.20){\special{em: lineto}}
\put(116.00,102.08){\special{em: lineto}}
\put(116.50,100.97){\special{em: lineto}}
\put(117.00,99.88){\special{em: lineto}}
\put(117.50,98.81){\special{em: lineto}}
\put(118.00,97.76){\special{em: lineto}}
\put(118.50,96.72){\special{em: lineto}}
\put(119.00,95.70){\special{em: lineto}}
\put(119.50,94.70){\special{em: lineto}}
\put(120.00,93.71){\special{em: lineto}}
\put(120.50,92.73){\special{em: lineto}}
\put(121.00,91.77){\special{em: lineto}}
\put(121.50,90.83){\special{em: lineto}}
\put(122.00,89.90){\special{em: lineto}}
\put(122.50,88.98){\special{em: lineto}}
\put(123.00,88.08){\special{em: lineto}}
\put(123.50,87.19){\special{em: lineto}}
\put(124.00,86.31){\special{em: lineto}}
\put(124.50,85.45){\special{em: lineto}}
\put(125.00,84.60){\special{em: lineto}}
\put(125.50,83.76){\special{em: lineto}}
\put(126.00,82.94){\special{em: lineto}}
\put(126.50,82.13){\special{em: lineto}}
\put(127.00,81.33){\special{em: lineto}}
\put(127.50,80.54){\special{em: lineto}}
\put(128.00,79.76){\special{em: lineto}}
\put(128.50,79.00){\special{em: lineto}}
\put(129.00,78.25){\special{em: lineto}}
\put(129.50,77.51){\special{em: lineto}}
\put(130.00,76.78){\special{em: lineto}}
\put(130.50,76.06){\special{em: lineto}}
\put(131.00,75.35){\special{em: lineto}}
\put(131.50,74.66){\special{em: lineto}}
\put(132.00,73.98){\special{em: lineto}}
\put(132.50,73.30){\special{em: lineto}}
\put(133.00,72.64){\special{em: lineto}}
\put(133.50,71.99){\special{em: lineto}}
\put(134.00,71.35){\special{em: lineto}}
\put(134.50,70.72){\special{em: lineto}}
\put(135.00,70.10){\special{em: lineto}}
\put(135.50,69.49){\special{em: lineto}}
\put(136.00,68.90){\special{em: lineto}}
\put(136.50,68.31){\special{em: lineto}}
\put(137.00,67.73){\special{em: lineto}}
\put(137.50,67.17){\special{em: lineto}}
\put(138.00,66.61){\special{em: lineto}}
\put(138.50,66.07){\special{em: lineto}}
\put(139.00,65.53){\special{em: lineto}}
\put(139.50,65.01){\special{em: lineto}}
\put(140.00,64.49){\special{em: lineto}}
\put(140.50,63.99){\special{em: lineto}}
\put(141.00,63.49){\special{em: lineto}}
\put(141.50,63.01){\special{em: lineto}}
\put(142.00,62.53){\special{em: lineto}}
\put(142.50,62.07){\special{em: lineto}}
\put(143.00,61.61){\special{em: lineto}}
\put(143.50,61.17){\special{em: lineto}}
\put(144.00,60.73){\special{em: lineto}}
\put(144.50,60.31){\special{em: lineto}}
\put(145.00,59.89){\special{em: lineto}}
\put(145.50,59.48){\special{em: lineto}}
\put(146.00,59.09){\special{em: lineto}}
\put(146.50,58.70){\special{em: lineto}}
\put(147.00,58.33){\special{em: lineto}}
\put(147.50,57.96){\special{em: lineto}}
\put(148.00,57.60){\special{em: lineto}}
\put(148.50,57.25){\special{em: lineto}}
\put(149.00,56.92){\special{em: lineto}}
\put(149.50,56.59){\special{em: lineto}}
\put(150.00,56.27){\special{em: lineto}}
\put(150.50,55.96){\special{em: lineto}}
\put(151.00,55.66){\special{em: lineto}}
\put(151.50,55.37){\special{em: lineto}}
\put(152.00,55.09){\special{em: lineto}}
\put(152.50,54.82){\special{em: lineto}}
\put(153.00,54.55){\special{em: lineto}}
\put(153.50,54.30){\special{em: lineto}}
\put(154.00,54.06){\special{em: lineto}}
\put(154.50,53.82){\special{em: lineto}}
\put(155.00,53.60){\special{em: lineto}}
\put(155.50,53.38){\special{em: lineto}}
\put(156.00,53.17){\special{em: lineto}}
\put(156.50,52.97){\special{em: lineto}}
\put(157.00,52.78){\special{em: lineto}}
\put(157.50,52.60){\special{em: lineto}}
\put(158.00,52.42){\special{em: lineto}}
\put(158.50,52.26){\special{em: lineto}}
\put(159.00,52.10){\special{em: lineto}}
\put(159.50,51.95){\special{em: lineto}}
\put(160.00,51.81){\special{em: lineto}}
\put(160.50,51.67){\special{em: lineto}}
\put(161.00,51.55){\special{em: lineto}}
\put(161.50,51.43){\special{em: lineto}}
\put(162.00,51.31){\special{em: lineto}}
\put(162.50,51.21){\special{em: lineto}}
\put(163.00,51.11){\special{em: lineto}}
\put(163.50,51.01){\special{em: lineto}}
\put(164.00,50.93){\special{em: lineto}}
\put(164.50,50.85){\special{em: lineto}}
\put(165.00,50.77){\special{em: lineto}}
\put(165.50,50.70){\special{em: lineto}}
\put(166.00,50.63){\special{em: lineto}}
\put(166.50,50.57){\special{em: lineto}}
\put(167.00,50.52){\special{em: lineto}}
\put(167.50,50.47){\special{em: lineto}}
\put(168.00,50.42){\special{em: lineto}}
\put(168.50,50.38){\special{em: lineto}}
\put(169.00,50.34){\special{em: lineto}}
\put(169.50,50.30){\special{em: lineto}}
\put(170.00,50.27){\special{em: lineto}}
\put(170.50,50.24){\special{em: lineto}}
\put(171.00,50.21){\special{em: lineto}}
\put(171.50,50.19){\special{em: lineto}}
\put(172.00,50.16){\special{em: lineto}}
\put(172.50,50.14){\special{em: lineto}}
\put(173.00,50.13){\special{em: lineto}}
\end{picture}

{\small \it
\begin{center}
Figure 4.
\\
The form of the size spectrum.
   \end{center}
}

\end{figure}

\subsubsection{Situation $h_{1} \geq 1, h_{2} \ll 1$}

The description of the process of  formation
of the droplets of the first sort  can't be simplified.
It has been already given in the previous sections.
But the process of formation of the droplets  of the second
sort
 is rather simple to describe.
The supersaturation is
determined  by  the vapor consumption  by the droplets
of the first sort. Then  one has the following expressions
\begin{equation} \label{73}
\theta_{2} = \exp ( - f_{*\ 2} \frac{n_{\infty}}{\eta_{tot\ 2}} \int_{0}^{z}
\exp ( - \Gamma_{2} \frac{ g_{1}  }
{ \Phi_{*} } ) dx ) \ \ ,
\end{equation}
\begin{equation} \label{74}
g_{2} = f_{*\ 2}  \int_{0}^{z} (z-x)^{3}
\exp ( -\Gamma_{2} \frac{ g_{1}  }
{ \Phi_{*} } )
\theta_{2} dx \ \ .
\end{equation}

The value of  $g_2$ during the period of the nucleation on the centers
of the second sort can be estimated as
$$
g_2 \ll \frac{\Phi_*}{\Gamma_2} \  \ ,
$$
which is based  on
$$
\delta_2 x  \ll \Delta_2 x             \ \ .
$$
So, $g_2$ is negligible.
It is necessary to calculate only $\theta_2$. To calculate $\theta_2$
one can get into account that the value of  $g_1$ grows so rapidly that
for the value  of
$ \int_0^z \exp(-\Gamma_2 \frac{g_1}{\Phi_*} )dx
$
one can show  the following approximation
$$
  \int_0^z \exp(-\Gamma_2 \frac{g_1}{\Phi_*}) dx
\approx
z \Theta(1-\frac{\Gamma_2 g_1}{\Phi_*}) +
\int_0^{\infty} \exp(-\Gamma_2 \frac{g_1}{\Phi_*}) dx
\Theta(\frac{\Gamma_2 g_1}{\Phi_*} - 1) \ \ ,
$$
or
$$
  \int_0^z \exp(-\Gamma_2 \frac{g_1}{\Phi_*}) dx
\approx
z \Theta(1-\frac{\Gamma_2 g_1}{\Phi_*}) +
z_b \Theta(\frac{\Gamma_2 g_1}{\Phi_*} - 1) \ \ ,
$$
where $z_b$ is extracted by the condition
$$ g_1(z_b) = \frac{\Phi_*}{\Gamma_2} $$
and $\Theta$ is the Heavisaid function.
The last approximation allows to calculate
$\theta_2$ analytically.

\subsubsection{Situation $h_{1} \geq 1, h_{2} \geq 1$ }

Actually, this situation
has been already analyzed in description of the situation
$\Delta_{1}x \ll \Delta_{2}x ; h_{1} \geq 1, h_{2} \ll 1$ .
Since $\Delta_{1}x \ll \Delta_{2}x$
one could not effectively use there the condition $h_{2} \ll 1$ because one
could not
assume  that  the inequality $h_{2} \ll 1$ ensures  the
pure exhaustion of  heterogeneous centers without any vapor exhaustion
and the consideration
made earlier couldn't been simplified.
As the result the previous consideration covers the situation
$h_{1} \geq 1, h_{2} \geq 1$.

\subsubsection{Situation $h_{1} \ll 1, h_{2} \ll 1$ }

From the first point of view it seems that
the situation $h_{1} \ll 1, h_{2} \ll 1$ has been  already described.
One has to stress that  $h_i \ll 1$ doesn't allow to state that the nucleation
process is going at the constant
 value of the supersaturation.
For the  first sort nucleation one
has the previous expressions (\ref{63}), (\ref{65}).
The analogous expressions
(\ref{63}), (\ref{65}) for the
 second sort nucleation can be violated.                                 So, the
process of nucleation  of the second sort droplets  can not
be described on the base of the initial value of the  supersaturation.

The calculation  of $g_{2}$ isn't necessary and only the calculation of
$\theta_{2}
$ is essential. One has
\begin{equation} \label{75}
\theta_{2\ (2)}= \exp(- f_{*\ 2}\frac{n_{\infty}}{\eta_{tot\ 2}} \int_{0}^{z}
\exp(( - (\frac{x}{\Delta_{h \ 1} x })^{3}) dx) \ \ ,
\end{equation}
 the final value for $\theta_{2}$ can be given by
\begin{equation} \label{76}
\theta_{2\ (2)}(\infty) = \exp[-f_{*\ 2} \Delta_{h\ 1} x
\frac{n_{\infty}}{\eta_{tot\ 2}}B]
\ \ .
\end{equation}
Eq. (\ref{75}) can be reduced by the simple rescaling to
\begin{equation} \label{75*}
\theta_{2\ (2)}= \exp(- a \int_{0}^{z}
\exp( - x^{3}) dx)
\end{equation}
with parameter $a =  f_{*\ 2} n_{\infty} \Delta_{h\ 1} x / \eta_{tot\ 2}
$. This behavior is drawn in Fig. 5.

\begin{figure}



{\small \it
\begin{center}
Figure 5.
\\
Function $\theta_{2\ (2)} (x) $ for different $a$.
   \end{center}
}

\end{figure}

The mono-disperse approximation
for the size spectrum of the first sort droplets
is based on the evident chain
of the inequalities
$$
\hat{\Delta} x_1 \sim \delta_1 x \ll \Delta_1 x \ll \Delta_2 x
\ \ . $$
When the mono-disperse approximation fails then the vapor consumption
isn't essential at all.

Now all imaginary  possible asymptotic situations have been studied. It doesn't mean
that all of them take place for two given sorts of heterogeneous centers
and some given  substance.

The second part of the manuscript which can be found in
the cond-mat e-print
archive at the same publishing date will complete the theory.

\end{document}